\documentclass[prc,twocolumn,showpacs,amsmath,amssymb]{revtex4}
\usepackage{graphicx}
\usepackage{dcolumn}
\usepackage{bm}
\begin{document}

\title{Coupled-channels density-matrix approach to low-energy nuclear collision dynamics: 
A technique for quantifying quantum decoherence effects on reaction observables}

\author{Alexis Diaz-Torres}

\affiliation{Department of Physics, University of Surrey, Guildford, GU2 7XH, 
United Kingdom}

\date{\today}

\begin{abstract}
The coupled-channels density-matrix technique for nuclear reaction dynamics, which is based on the Liouville-von Neumann equation with Lindblad dissipative terms, is developed with the inclusion of full angular momentum couplings. It allows a quantitative study of the role and importance of quantum decoherence in nuclear scattering. Formulae of asymptotic observables that can reveal effects of quantum decoherence are given. A method for extracting energy-resolved scattering information from the time-dependent density matrix is introduced. As an example, model calculations are carried out for the low-energy collision of the $^{16}$O projectile on the $^{154}$Sm target. 
\end{abstract}


\maketitle

\section{Introduction} 

Low-energy nuclear reaction dynamics has successfully been treated within the stationary-state multi-channel scattering theory including complex potentials \cite{Feshbach,Taylor}. 
However, this \emph{cannot} account for quantum decoherence \cite{Zeh0,Zurek,Schlosshauer}, which is a key aspect of irreversibility in open dynamical systems \cite{Breuer}, when unavoidably a limited number of degrees of freedom and reaction channels is used \cite{Alexis0}. This issue can be tackled through the present time-dependent approach. The coupled-channels density-matrix ({\sc ccdm}) technique was first introduced in studies of quantum molecular dynamics \cite{Saalfrank}, and has recently been applied to investigate the coupling-assisted quantum tunneling in heavy-ion fusion \cite{Alexis1,Alexis2}. Ref. \cite{Alexis3} provides a didactic discussion on the {\sc ccdm} approach. In contradistinction to this approach, most of the dynamical models \cite{Alexis1} of dissipative nuclear collisions do \emph{not} treat the relative motion of the nuclei quantum-mechanically and/or use \emph{incoherent} (statistically averaged) rather than \emph{decoherent} (partially coherent) reaction channels.  

\begin{figure}
\includegraphics[width=0.50\textwidth,angle=0]{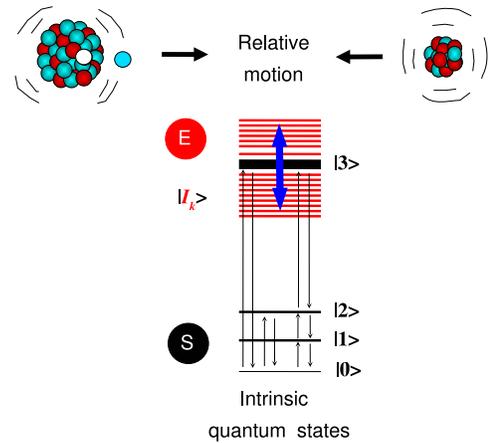}
\caption{A low-energy nuclear collision represented by an open quantum system (relative motion + a few intrinsic, low-lying collective states). The bath of single-particle states surrounding a giant resonance state represents the environment. It gradually destroys the coherent quantum superposition of the reduced-system collective states, as the nuclei approach.}
\label{Fig1}
\end{figure}

Figure \ref{Fig1} illustrates the innovative ideas of the {\sc ccdm} approach. The situation can be imagined as an orchestra (representing the reduced quantum system), where the director plays the role of the relative motion and the musicians correspond to a few intrinsic, low-lying collective states. Now imagine an airplane appearing overhead, representing the environment of innumerable nucleonic states. When the airplane is not present, the orchestra plays a marvelous music, all musicians are in sync, like in a coherent quantum superposition. But when the airplane approaches, the listener percieves two effects due to interference: (i) the music gets attenuated (dissipation) and, most importantly, (ii) the quality of the music changes, as the musicians play out of sync (decoherence).  
 
Decoherence, which always accompanies dissipation in open quantum systems \cite{Breuer}, means dynamical dislocalization of coherent quantum superpositions due to entanglement of the system with its environment \cite{Zeh1}. Coherent quantum superpositions are the basis of the coherent coupled-channels approach to near-barrier reaction dynamics, which manifest themselves through experimental fusion-barrier distributions \cite{Nanda0}. Dissipation of energy and angular momentum is revealed in heavy-ion deep-inelastic scattering that occurs at near-barrier energies as well \cite{Hinde1}. Cross sections of deep-inelastic collisions also indicate loss of angular-momentum coherence \cite{Simbel1,Rosa1}. While the coherent coupled-channels calculations \cite{Balantekin1} are able to explain several collision observables, major problems are unresolved. Foremost is the inability to describe the elastic and quasi-elastic scattering and fusion processes simultaneously \cite{Newton1} and the related, more recent failure to describe consistently below-barrier quantum tunneling and above-barrier fusion yields \cite{Nanda1}. New, precise fusion measurements have inevitably led to phenomenological (sometimes contradictory) adjustments \cite{Rae,Esbensen1,Hagino1} to stationary-state coupled-channels models to fit the experimental data, but without a physically consistent foundation. Complementary to fusion experiments, precision measurements of back-angle scattering energy spectra \cite{Gasques1,Evers1,Lin1} and quasi-elastic barrier distributions \cite{Piasecki1} clearly indicate that complex, dynamical processes (other than the low-lying collective excitation of the interacting nuclei \cite{Yusa}) play an important role in low-energy heavy-ion collisions. 

This paper suggests that quantum decoherence and dissipation should be simultaneously included in a consistent description of low-energy reaction dynamics, when a restricted set of (relevant) degrees of freedom is employed. A possible description is the {\sc ccdm} approach whose formalism is described in Sec. II. For the sake of simplicity and a specific application, without loss of generality, the collision of an inert spherical $^{16}$O projectile on a deformed $^{154}$Sm target is studied. It includes the couplings between the relative orbital angular momentum $\hat{L}$ of the reactants and the spin $\hat{I}$ of the ground-state rotational-band levels of $^{154}$Sm. Formulae of asymptotic observables that can reveal effects of quantum decoherence are obtained. A method for calculating the energy-resolved observables is presented. In Sec. III, model calculations are carried out and discussed, while a summary is given in Sec. IV.  

\section{Formalism}
\subsection{Initial density matrix}

The origin of the reference frame is in the overall center of mass. The vector $\vec{r}$ refers to the relative vector between the projectile and the target, while $\hat{r}$ and $\hat{\xi}$ are respectively the orientation angles of $\vec{r}$ and the target intrinsic symmetry axis, with respect to the laboratory fixed z-axis defined by the beam direction. 

Initially, the target is considered to be unpolarized at the ground-state of spin $I_0$, being the rotational state the $Y_{I_0 M_{I_0}}(\hat{\xi})$ spherical harmonic. It is coupled to the orbital motion described with $Y_{L M_L}(\hat{r})$. The total 
angular-momentum state, $|I_0 L;JM \rangle = \sum_{M_L M_{I_0}} C_{LM_L I_0 M_{I_0}}^{JM}Y_{LM_L}(\hat{r})Y_{I_0M_{I_0}}(\hat{\xi)}$, describes the angular variables, denoting $C$ the Clebsch-Gordan coefficients. The radial motion is described with a Gaussian $\psi_{{k}_0}(r)$ wave-packet, centered at $r_0$ with the average radial wave-number $k_0$ and the spatial dispersion $\sigma$:
\begin{equation}
\psi_{k_0}(r) = N \exp \, [-\frac{(r-r_0)^2}{2\sigma^2} ] \, 
                              e^{ik_0 \, r}, 
\label{eq1}
\end{equation}
where the constant $N$ is specified below. Thus, the initial state is $|\chi \rangle = \, \sum_{LJM}\psi_{k_0}(r) \, |I_0 L;JM \rangle$, and the initial density operator is $\hat{\rho}_0=(2I_0+1)^{-1} \, |\chi \rangle \langle \chi |$. The latter can be represented as 

\begin{equation}
\hat{\rho}_0 \, = \sum_{\alpha,\alpha ',rs} \, |r ) |\alpha \rangle \, \rho_{\alpha \alpha '}^{rs}(t=0) \,
\langle \alpha '| ( s|,  \label{eq2}
\end{equation}
where $\alpha \equiv (IL;JM)$, $|\alpha \rangle$ and $|r )$ are the coupled angular-momentum basis and the discrete grid-basis describing the internuclear separations, respectively. The initial density matrix is: 

\begin{eqnarray}
\rho_{\alpha \alpha '}^{rs}(t=0)&=& (2I_0 + 1)^{-1} N^2 \exp \, [-\frac{(r-r_0)^2}{2\sigma^2} ] \, e^{ik_0 \, r} 
 \nonumber \\ 
&& \times \exp \, [-\frac{(s-r_0)^2}{2\sigma^2} ] e^{-ik_0 \, s} \,\, \delta_{II_0} \,\, \delta_{I'I_0},
\label{eq3}
\end{eqnarray}
where $N$ is determined from the normalization condition $\sum_{r \alpha} \rho_{\alpha \alpha}^{rr} = 1$.

\subsection{Time evolution}
\subsubsection{Non-dissipative case}

The Liouville-von Neumann master equation dictates the time evolution of the density matrix operator $\hat{\rho}(t)$ with the initial value (\ref{eq2}). It reads as
\begin{equation}
i\hbar \, \frac{\partial \hat{\rho}}{\partial t} = [\hat{H}, \hat{\rho}],
\label{eq4}
\end{equation}
where $\hat{H}$ is the system (Hermitian) Hamiltonian specified below. Inserting the expansion (\ref{eq2}) for $\hat{\rho}(t)$ in (\ref{eq4}), and projecting onto the basis states, we get a system of coupled equations for the matrix elements $\rho_{\alpha \alpha '}^{rs}(t)$:
\begin{equation}
i\hbar \, \dot{\rho}_{\alpha \alpha '}^{rs} = \sum_{t\beta} \, ( \, H_{\alpha \beta}^{rt} \, 
\rho_{\beta \alpha '}^{ts} \, - \, \rho_{\alpha \beta}^{rt} \, H_{\beta \alpha '}^{ts} \, ), 
\label{eq5}
\end{equation}
with the initial values (\ref{eq3}). The system Hamiltonian contains different pieces: (i) the radial kinetic energy operator $\hat{T}$, (ii) the total bare (Coulomb+nuclear+centrifugal) nucleus-nucleus real potential $\hat{U}$, (iii) the total (Coulomb+nuclear) real coupling potential $\hat{V}$ between the relative motion and the intrinsic rotational states of the target, and (iv) the target intrinsic Hamiltonian $\hat{H_0}$. In terms of these operators, the coupled equations (\ref{eq5}) read as
\begin{eqnarray}
i\hbar \, \dot{\rho}_{\alpha \alpha '}^{rs}&=&\sum_{t}\, (\, T^{rt} \, 
\rho_{\alpha \alpha '}^{ts} \, - \, \rho_{\alpha \alpha '}^{rt} \, T^{ts}\, ) \nonumber \\
&& + \, [\, U_{\alpha} (r) \, - \, U_{\alpha '} (s)\, ]\, \rho_{\alpha \alpha '}^{rs} \nonumber \\
&& + \, \sum_{\beta}\, [\, V_{\alpha \beta}(r) \, 
\rho_{\beta \alpha '}^{rs} \, - \, \rho_{\alpha \beta}^{rs} \, V_{\beta \alpha '} (s)\, ] \nonumber \\
&& + \, (\, \varepsilon_{\alpha} \, - \, \varepsilon_{\alpha '} \, )\, \rho_{\alpha \alpha '}^{rs} . 
\label{eq6}
\end{eqnarray}

\subsubsection{Dissipative case}

In practice, however, a truncated model space of reaction channels (\emph{reduced system}) is employed \cite{Amos}. The impact of the excluded model space (\emph{environment}) on the reaction dynamics is usually treated through complex potentials \cite{Feshbach,Taylor,Amos}, making the system Hamiltonian non-Hermitian. This results in loss of probability and energy from the reduced system to the environment, \emph{but} the quantum coherence in the basis states of the reduced system is preserved, as recently demonstrated in Ref. \cite{Alexis0}. It is inconsistent with the irreversible dynamics of an open quantum system \cite{Breuer}, as energy dissipation always goes hand in hand with quantum decoherence \cite{Zeh0,Zurek,Schlosshauer}. 

Irreversibility can be consistently described by the Lindblad master equation \cite{Lindblad1,Gorini,Sandulescu1,Ronnie1}. Here, quantum decoherence and dissipation are incorporated into the dynamics through a dissipative Liouvillian, 
$i \hbar \mathcal{L}_D\, \hat{\rho}$, which is added to the r.h.s. of Eq. (\ref{eq4}). 
$\mathcal{L}_D\, \hat{\rho}$ reads as 
\begin{equation} 
\mathcal{L}_D \, \hat{\rho} \, = \, \sum_\nu \bigl( \hat{\mathcal C}_\nu \, \hat{\rho} \, \hat{\mathcal C}_\nu^{\dag} -
\frac{1}{2} \bigl[\hat{\mathcal C}_\nu^{\dag} \, \hat{\mathcal
C}_\nu ,\hat{\rho} \bigl]_{+} \bigl)\, 
\label{eq6a}
\end{equation}
where $[\ldots]$ and $[\ldots]_{+}$ denote the commutator and anti-commutator, respectively. Each $\hat{\mathcal C}_\nu$ is a Lindblad operator for a 
dissipative coupling, physically motivated according to the specific problem. 
It is assumed that each dissipative coupling $\nu \equiv (\alpha \alpha')$ between 
given states $|\alpha \rangle$ and $|\alpha' \rangle$ has an
associated (local) radial-dependent decay rate $\Gamma_{\alpha \alpha'}^{r r}$ \cite{Saalfrank2}, i.e., the \emph{spontaneous emission} Lindblad operator $\hat{\mathcal C}_{\alpha \alpha'}=\sqrt{\Gamma_{\alpha \alpha'}^{rr}} |\alpha \rangle \langle \alpha'|$.

In order to also describe decay to the Hilbert space of excluded, intrinsic degrees of freedom, environmental states \cite{Irene,Bertlmann} are considered in addition to the basis $|\alpha \rangle$ of the reduced system. All these states are assumed to be orthonormal, resulting in an enlarged basis $|\bar{\alpha} \rangle$. Using this and the discrete grid-basis, the matrix elements of (\ref{eq6a}) read as \cite{Erratum} 
\begin{eqnarray}
(\mathcal{L}_D\, \hat{\rho})_{\bar{\alpha} \bar{\alpha}'}^{r s} &=&
\delta_{\bar{\alpha} \bar{\alpha}'}\, \sum_{\mu} \sqrt{\Gamma_{\bar{\alpha} \mu}^{rr}} \, 
\rho_{\mu \mu}^{rs} \, \sqrt{\Gamma_{\bar{\alpha} \mu}^{ss}} \, \nonumber \\ 
&& - \, \frac{1}{2}\, \sum_{\mu}\, ( \, \Gamma_{\mu \bar{\alpha}}^{rr} \, + \, 
\Gamma_{\mu \bar{\alpha}'}^{ss} \,)\, 
\rho_{\bar{\alpha} \bar{\alpha}'}^{rs},
\label{eq6b}
\end{eqnarray}
where $\mu$ runs over all the $\bar{\alpha}$ states, and 
$\Gamma_{\bar{\alpha} \bar{\alpha}}^{rr} = \sum_{\mu \ne \bar{\alpha}} \Gamma_{\mu \bar{\alpha}}^{rr}$ \cite{Saalfrank3}.

We now distinguish two kinds of density matrix elements $\rho_{\bar{\alpha} \bar{\alpha}'}^{r s}$, namely one type associated with the reduced-system states $|\alpha \rangle$ only, and another type related to the environmental states 
$|\bar{\alpha} \rangle, \, \bar{\alpha} \ne  \alpha$. These obey the following equations of motion: 
\begin{eqnarray}
i\hbar \, \dot{\rho}_{\alpha \alpha '}^{rs}&=&\sum_{t}\, (\, T^{rt} \, 
\rho_{\alpha \alpha '}^{ts} \, - \, \rho_{\alpha \alpha '}^{rt} \, T^{ts}\, ) \nonumber \\
&& + \, [\, U_{\alpha} (r) \, - \, U_{\alpha '} (s)\, ]\, \rho_{\alpha \alpha '}^{rs} \nonumber \\
&& + \, \sum_{\beta}\, [\, V_{\alpha \beta}(r) \, 
\rho_{\beta \alpha '}^{rs} \, - \, \rho_{\alpha \beta}^{rs} \, V_{\beta \alpha '} (s)\, ] \nonumber \\
&& + \, (\, \varepsilon_{\alpha} \, - \, \varepsilon_{\alpha '} \, )\, \rho_{\alpha \alpha '}^{rs}  \nonumber \\
&& + \, i\hbar \, \, \{ \, \delta_{\alpha \alpha'}\, \sum_{\mu} \sqrt{\Gamma_{\alpha \mu}^{rr}} \, 
\rho_{\mu \mu}^{rs} \, \sqrt{\Gamma_{\alpha \mu}^{ss}} \, \nonumber \\ 
&& - \, \frac{1}{2}\, \sum_{\mu}\, ( \, \Gamma_{\mu \alpha}^{rr} \, + \, 
\Gamma_{\mu \alpha'}^{ss} \,)\, \rho_{\alpha \alpha'}^{rs} \, \}, 
\label{eq6c}
\end{eqnarray}
for matrix elements of the reduced-system states, whereas 
\begin{eqnarray}
\dot{\rho}_{\bar{\alpha} \bar{\alpha}'}^{rs}&=&
\delta_{\bar{\alpha} \bar{\alpha}'}\, \sum_{\mu} \sqrt{\Gamma_{\bar{\alpha} \mu}^{rr}} \, 
\rho_{\mu \mu}^{rs} \, \sqrt{\Gamma_{\bar{\alpha} \mu}^{ss}} \, \nonumber \\ 
&& - \, \frac{1}{2}\, \sum_{\mu}\, ( \, \Gamma_{\mu \bar{\alpha}}^{rr} \, + \, 
\Gamma_{\mu \bar{\alpha}'}^{ss} \,)\, \rho_{\bar{\alpha} \bar{\alpha}'}^{rs}, 
\label{eq6d} 
 \end{eqnarray}
for matrix elements involving the environmental states, i.e., either $\bar{\alpha}$ or 
$\bar{\alpha}' \ne \alpha$, at least. It is worth emphasizing that the environmental states are not reaction channels, but auxiliary states \cite{Irene} supplying a probability drain and rendering the reduced-system states decoherent. 

The initial values for Eqs. (\ref{eq6c}) are given by (\ref{eq3}), while for Eqs. (\ref{eq6d}) these are zero as the nuclei are initially (far apart) at the ground-states. Hence, the off-diagonal elements $\rho_{\bar{\alpha} \bar{\alpha}'}^{rs}$ in (\ref{eq6d}) remain zero. However, the diagonal terms $\rho_{\bar{\alpha} \bar{\alpha}}^{rs}$ absorb probabilities only, provided there is no flux back from the environment to the reduced system. 

In the calculations below, it will be considered that the off-diagonal elements of the decay-rate matrix $\Gamma$ are \emph{nonzero} only for transitions from the reduced system to the environment, i.e., $\Gamma_{\bar{\alpha} \alpha}^{rr}, \, \bar{\alpha} \ne \alpha$. 
(Still, environment-induced transitions among the states of the reduced system may occur.) The decay rates are given by $\Gamma_{\bar{\alpha} \alpha}^{rr} = W_{\alpha}(r) /\hbar$, where $W_{\alpha}(r) > 0$ are decay functions. These are here treated as \emph{empirical} functions, but it is hoped that a microscopic theory on damping of collective excited states \cite{Matsuo,Guo,Ring}, including dynamical modifications of excited state properties due to the close proximity of other nucleus \cite{McIntosh}, can provide them. 

Various types of environments can coexist in a nuclear collision, which may be specific to particular degrees of freedom, such as isospin asymmetry or weak binding. Among these environments, which can be coupled to specific states or to all states of the reduced system, are (i) the high level-density of one- and multi-nucleonic excitations in different mass/charge partitions (transfer), (ii) the continuum of non-resonant decay states of weakly-bound nuclei (breakup), and (iii) the innumerable nuclear molecular (compound nucleus) states (fusion). These can be treated separately, and their effects can be distinguished within the {\sc ccdm} approach.

\subsection{Asymptotic observables}

Having the solution of Eqs. (\ref{eq6c}), after a long period of time $t_f$, when the centroid of the recoiled body of the wave-packet is at a large internuclear distance and a quasi-stationary probability current establishes in all the $\alpha$ channels, we then calculate asymptotic observables. For instance, these can be the angular distribution of the target excitations, and their integrated values. These are calculated as follows.

We now introduce the projector $\hat{P}_{I M_I} = |I M_I \rangle \langle I M_I |$ associated with a specific state of the target. The new operator $\hat{\rho}(t_f)\,\hat{P}_{I M_I}$ describes the scatterred waves in this target state
\begin{eqnarray}
\hat{\rho}(t_f)\,\hat{P}_{I M_I}&=&\langle I M_I |\hat{\rho}(t_f)|I M_I \rangle 
= \nonumber \\
&& \sum_{p} \, C_{Lm I M_I}^{JM} \, Y_{Lm}(\hat{r}) \, \, |r ) \, \,
\rho_{\gamma \lambda }^{rs}(t_f) \nonumber \\ 
&& \times \, ( s| \, \, C_{L'm' I M_I}^{J'M'} \, Y_{L'm'}^*(\hat{s}),
\label{eq7}
\end{eqnarray}       
where $p \equiv (r,s,L,m,J,M,L',m',J',M')$, $\gamma \equiv (IL;JM)$ and 
$\lambda \equiv (IL';J'M')$. The reduced density matrix, $\rho_{\gamma \lambda }^{rs}(t_f)$, is normalized with its trace, i.e., $\sum_{r \gamma} \rho_{\gamma \gamma}^{rr}(t_f)$.  

We note that (\ref{eq7}) is still an operator in $\vec{r}$ and $\vec{s}$. The radial projector $\hat{P}_{r'} = |r') \, ( r'|$ associated with a specific separation between the nuclei is then introduced. With this projector we now act on (\ref{eq7}) and set 
$\hat{r}=\hat{s}=\hat{r}'$. A new operator is thus obtained, whose partial trace (sum over all separations $r'$) provides the probability for producing the target in state ($I,M_I$) with the relative coordinate in the direction $\hat{r}'$:
\begin{eqnarray}
\frac{d\mathcal{W}}{d\Omega}(I,M_I)&=&\sum_{q} \, C_{Lm I M_I}^{JM} \, Y_{Lm}(\hat{r}') \, 
\mathcal{S}_{\gamma \lambda } (t_f) \nonumber \\
&& \times \, C_{L'm' I M_I}^{J'M'} \, Y_{L'm'}^*(\hat{r}'),
\label{eq8}
\end{eqnarray}  
where $q \equiv (L,m,J,M,L',m',J',M')$ and $\mathcal{S}_{\gamma \lambda }(t_f) = \sum_{r'} \rho_{\gamma \lambda }^{r'r'}(t_f)$. The latter contains information about the coherence of angular momenta.

Integrating (\ref{eq8}) over all directions $\hat{r}'$ of solid angles, and summing over all $M_I$, the total probability for producing the target in state $I$ (population) is obtained:
\begin{equation}
\mathcal{W}(I) = \sum_{M_I} \, \sum_{LmJM} \, (\, C_{Lm I M_I}^{JM} \,)^2 \, 
\mathcal{S}_{\gamma \gamma } (t_f).
\label{eq9}
\end{equation}

\subsection{Energy-resolved observables} 

The observables (\ref{eq8}) and (\ref{eq9}) correspond to average values for the range of energies contained in the incident wave-packet (\ref{eq1}). The energy-resolved scattering information can be obtained using a window operator \cite{Schafer}. 
The key idea is to calculate, for definite $\gamma \lambda$ indices, the energy spectrum 
$\mathcal{P}(E_k)$ of the initial and final reduced density matrices. $E_k$ is the centroid of a \emph{total energy bin} of width $2\epsilon$. A matrix of reflection coefficients, 
$\mathcal{R}_{\gamma \lambda } (E_k)$, is determined by the ratio
\begin{equation}
\mathcal{R}_{\gamma \lambda } (E_k)\, = \, 
\frac{\mathcal{P}_{\gamma \lambda}^{\, final}(E_k)}
{\sum_{\gamma} \, \mathcal{P}_{\gamma \gamma}^{\, initial}(E_k)}, 
\label{eq10a}
\end{equation}
which replaces the matrix $\mathcal{S}_{\gamma \lambda } (t_f)$ in (\ref{eq8}) and (\ref{eq9}). Expression (\ref{eq10a}) generalizes the wave-packet formulation of the reflection coefficient \cite{Tannor,Yabana}.  

The energy spectrum $\mathcal{P}(E_k)\,=\,\widetilde{\textnormal{Tr}}$($\hat{\Delta}\hat{\rho}$), where $\widetilde{\textnormal{Tr}}$ denotes a partial trace involving the radial indices only, and $\hat{\Delta}$ is the window operator \cite{Schafer}:
\begin{equation}
\hat{\Delta}(E_k,n,\epsilon)\, \equiv \, \frac{\epsilon^{2^n}}  
{[(\hat{\mathcal{H}}\, - \, E_k)^{2^n}\, + \, \epsilon^{2^n}]}, 
\label{eq10b}
\end{equation}      
where $\hat{\mathcal{H}}$ is the system asymptotic Hamiltonian, and $n$ determines the shape of the window function. As $n$ is increased, this shape rapidly becomes rectangular with very little overlap between adjacent energy bins \cite{Schafer}, remaining the bin width constant at $2\epsilon$. The spectrum is constructed for a set of $E_k$ where $E_{k+1}=E_k + 2\epsilon$. Thus, scattering information over a range of incident energies can be extracted from a time-dependent density matrix that has been calculated on a grid.

\subsubsection{Example}

\begin{figure}
\begin{center}
\includegraphics[width=0.5\textwidth,angle=0]{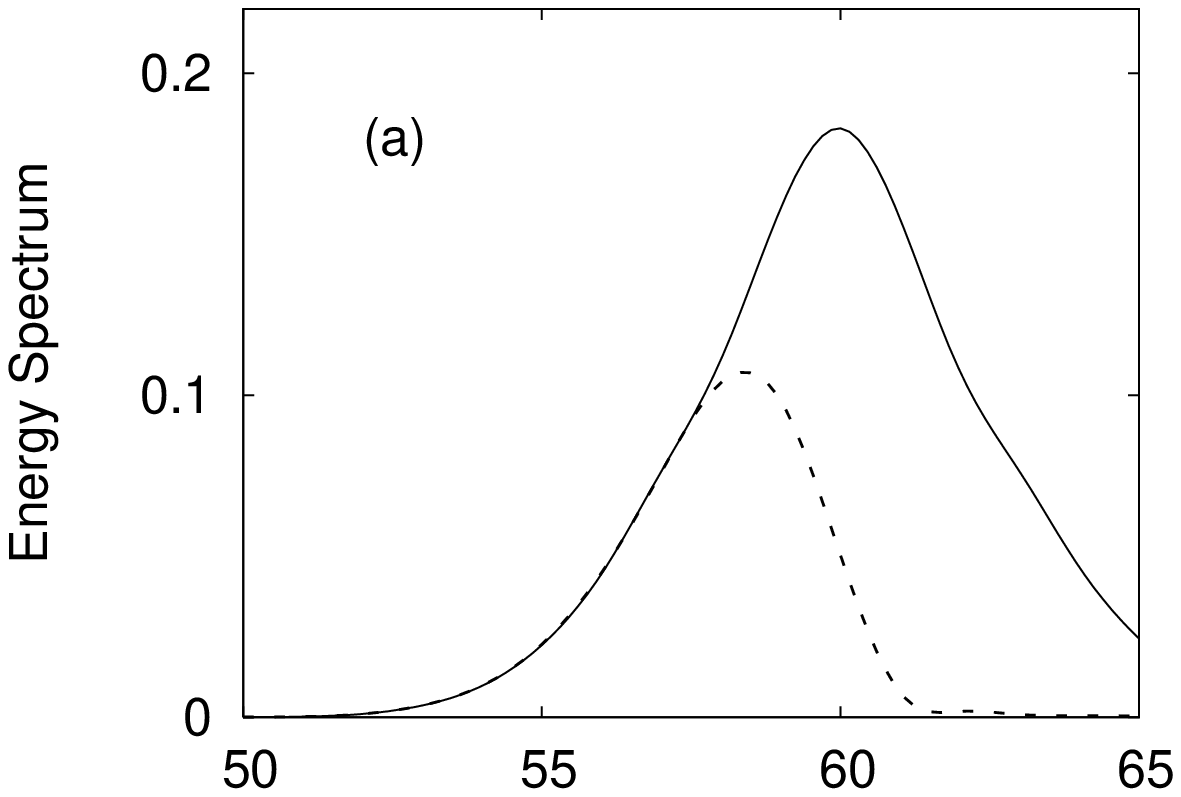} \\
\includegraphics[width=0.5\textwidth,angle=0]{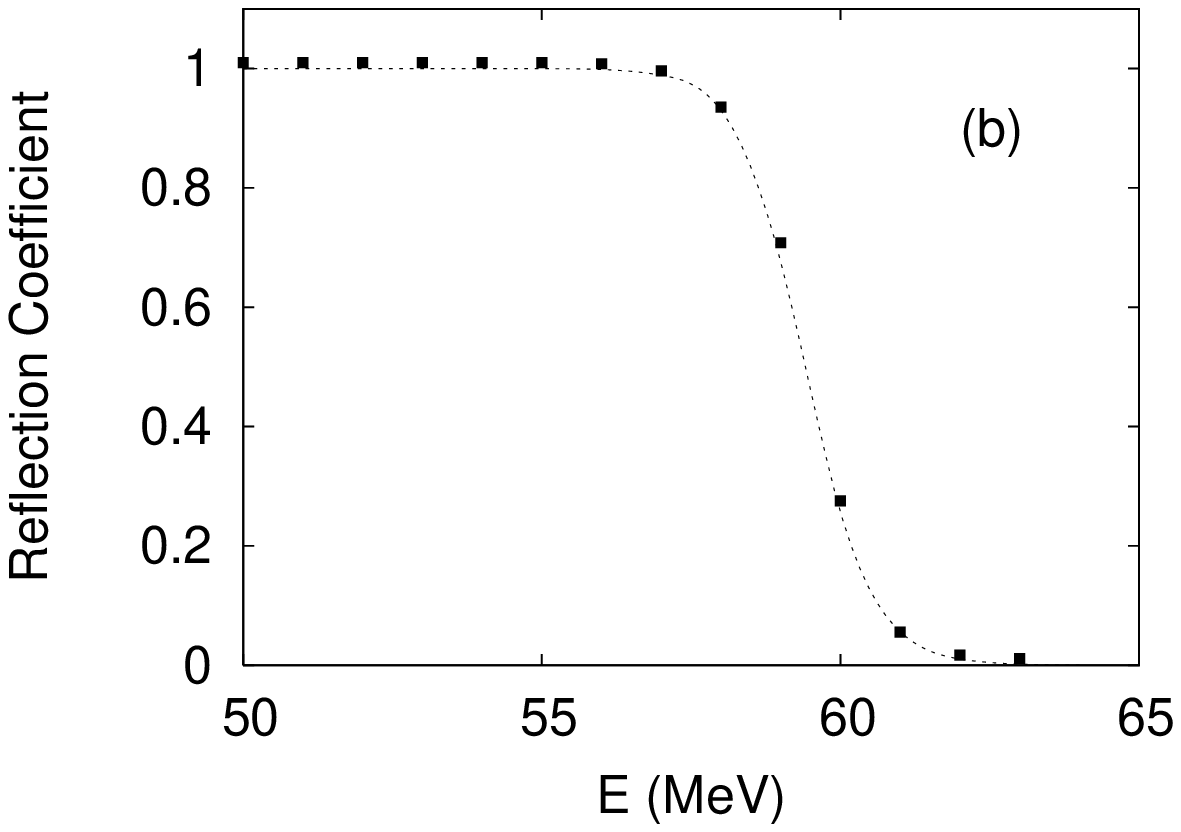}
\caption{(a) Energy spectrum of the initial (solid line) and final (dashed line) density matrices for the elastic scattering of $^{16}$O off the $^{154}$Sm target at the average total energy $E_0 = 60$ MeV. (b) Energy-resolved reflection coefficients provided by the final-to-initial ratio of the energy spectrum (full squares). These very well agree with those of a stationary Schr\"odinger equation (dotted line).}
\label{Fig2}
\end{center}
\end{figure}

Figure \ref{Fig2} shows for a single-channel {\sc ccdm} calculation with 
$L=0$ \cite{Alexis1}: (a) the energy spectrum ($n=4$, $2\epsilon=1$ MeV) of the initial (solid line) and final (dashed line) density matrices, and (b) the final-to-initial ratio of the energy spectrum providing energy-resolved reflection coefficients (full squares). These very well agree with those (dotted line) of a time-independent Schr\"odinger equation with a short-range imaginary potential or an ingoing-wave boundary condition at small radii, as implemented in the {\sc ccfull} code \cite{CCFULL}.
 
In the {\sc ccdm} calculation above, the grid ($r=0-100$ fm) was evenly spaced with $512$ radial points. The incident wave-packet was initially centered at $r_0 = 50$ fm, with width $\sigma = 5$ fm, and was boosted toward the potential with the appropriate average kinetic energy for the total energy $E_0 = 60$ MeV. 

The time evolution of the density matrix was carried out using a Faber propagator \cite{Faber} and the Fourier method \cite{Kosloff2} for the commutator between the kinetic energy and density operator. The time step for the density-matrix propagation was $\Delta t = 10^{-22}$ s.

The form of the bare nuclear potential between the $^{16}$O and $^{154}$Sm nuclei is a Woods-Saxon potential with ($V_0, r_0,a_0) \equiv$ ($-$165 MeV, 0.95 fm, 1.05 fm). The Coulomb potential is that for two point charges. These yield a $s$-wave Coulomb barrier of $V_B = 59.41$ MeV at the radius $R_B = 10.81$ fm. 

The irreversible capture of the two nuclei in the nucleus-nucleus potential pocket inside the Coulomb barrier (fusion) is caused by an environmental coupling \cite{Alexis1,Alexis2} with a decay function $W(r)$ taken as a Fermi function with depth $10$ MeV and diffuseness $0.1$ fm, located at the pocket radius of $5.77$ fm.

\section{Model calculations}

In the model calculations the $^{16}$O projectile was taken to be inert and the $^{154}$Sm target was allowed to be excited up to the $4^{+}$ state of the ground-state rotational band. The all order nuclear coupling of the ground-state $0^{+}$ to the states $2^{+}$ and $4^{+}$, with excitation energies $E_{2^+}=0.08$ MeV and $E_{4^+}=0.27$ MeV, has a macroscopic deformed Woods-Saxon form with the radius parameter of $1.06$ fm, and the deformation parameters of 
$\beta_2=0.322$ and $\beta_4=0.027$. The Coulomb coupling includes terms up to second order with respect to $\beta_2$ and to the first order of $\beta_4$. The total coupling-potential matrix in the coupled angular-momentum basis is calculated using the {\sc ccfull} code \cite{CCFULL} and the Kermode-Rowley matrix technique \cite{KR}.

With such a coupling Hamiltonian, the time propagation on the grid employed in Fig. \ref{Fig2} is very time consuming and memory demanding, beyond the present limit of our computational capability. That is why the present calculations aim at exploring \emph{qualitative} effects only, for which a smaller grid ($r=0-40$ fm) with $64$ evenly spaced radial points suffices. The initial wave-packet is then centered at $r_0 = 25$ fm, with width $\sigma = 3$ fm and the incident, average total energy $E_0 = 60$ MeV. Relative partial waves up to $20 \hbar$ are included in the calculations.

\begin{figure}
\begin{center}
\includegraphics[width=0.5\textwidth,angle=0]{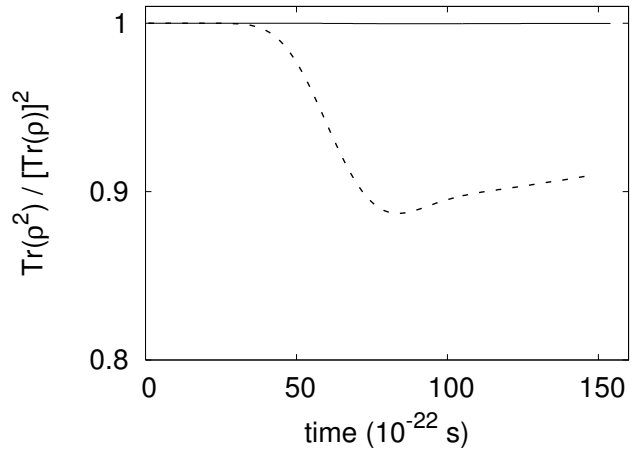}
\caption{Time evolution of the measure of coherence in the reduced system. While the fusion environment essentially preserves coherence (solid line), it is destroyed by the surface environment (dashed line).}
\label{Fig3}
\end{center}
\end{figure}

Two types of calculations (without energy projection) are carried out including: (i) only the effects of the \emph{fusion} environment highlighted above, and (ii) in addition the effects of a \emph{surface} environment specified below. While the fusion environment is coupled to all the $^{154}$Sm states, the surface environment is considered to be coupled to the ground-state only. The latter can be associated with complex, multi-nucleon/cluster transfers from the ground-state of the colliding nuclei to other mass (or charge) partitions. The corresponding decay function is taken as a Gaussian with width of $1$ fm, centered at the contact radius that is estimated as $1.2(16^{1/3} + 154^{1/3})$ fm. This function is physically motivated by the spatial localization of transfer processes in heavy-ion reactions \cite{Satchler}. The measure of coherence \cite{Alexis0,Klum} in the reduced system is the ratio Tr$(\hat{\rho}^2)/[$Tr$(\hat{\rho})]^2$, whose time evolution is presented in Fig. \ref{Fig3}. The fusion environment essentially preserves coherence (solid line), while the surface environment results in decoherence (dashed line). Comparing the calculation (ii) to (i), we learn how the surface environment-induced decoherence impacts on the asymptotic observables (\ref{eq8}) and (\ref{eq9}). 

\begin{figure}
\begin{center}
\includegraphics[width=0.5\textwidth,angle=0]{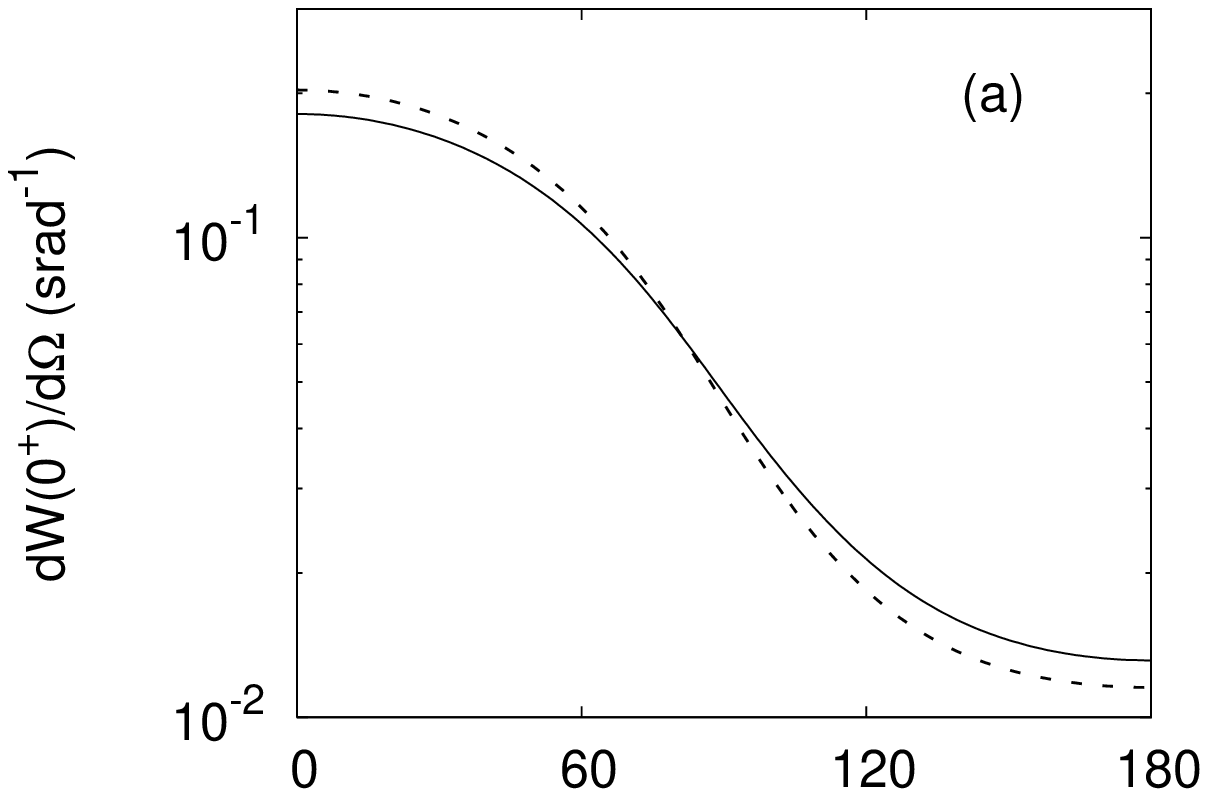} \\
\includegraphics[width=0.5\textwidth,angle=0]{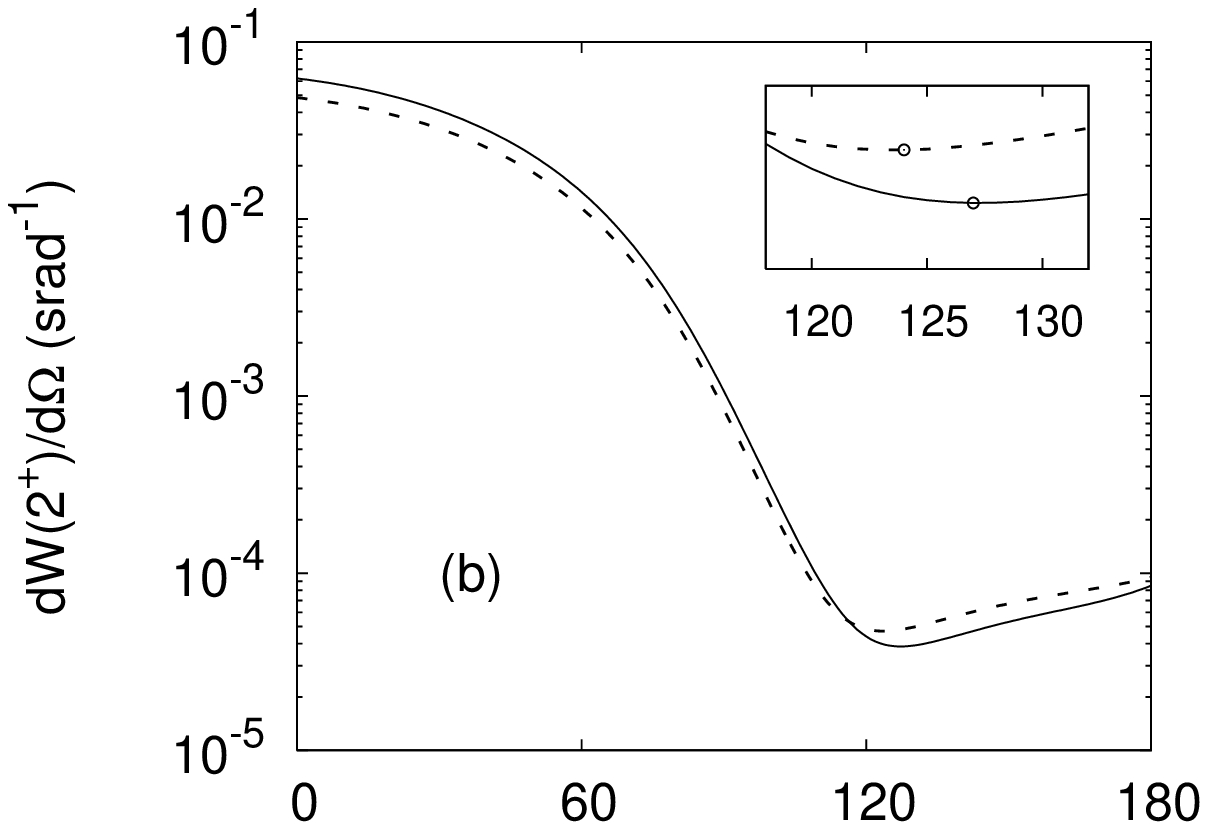} \\
\includegraphics[width=0.5\textwidth,angle=0]{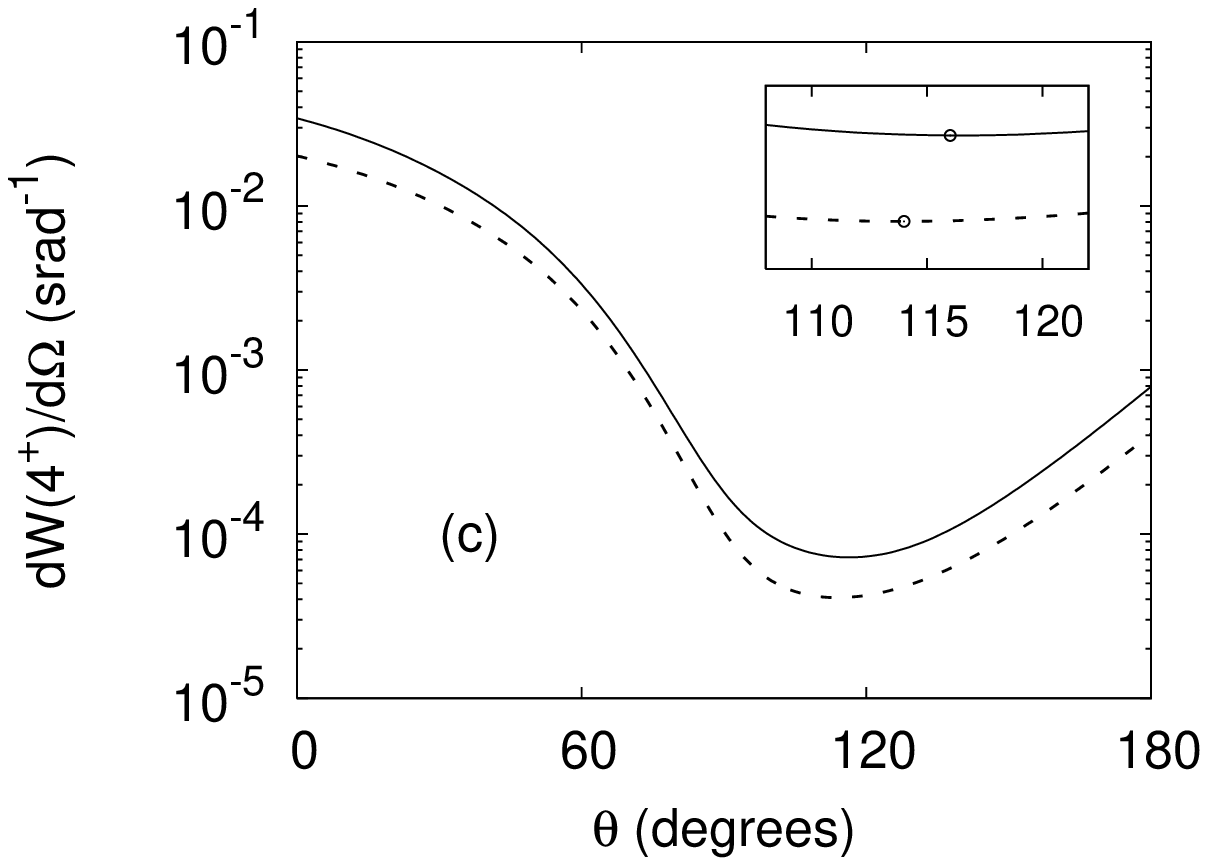}
\caption{Angular distribution of the $^{154}$Sm excitation probabilities, when an inert $^{16}$O projectile scatters off the $^{154}$Sm target at the average total energy $E_0 = 60$ MeV: (a) $0^{+}$, (b) $2^{+}$, and (c) $4^{+}$. The solid lines are outcomes including the effects of the fusion environment only, while the dashed lines include in addition the effects of the surface environment. The latter destroys the coherence of relative partial waves, changing by a few degrees the minimum of the inelastic distributions (small figures inserted).}
\label{Fig4}
\end{center}
\end{figure}

Figure \ref{Fig4} shows the angular distribution of the $^{154}$Sm excitations, which corresponds to expression (\ref{eq8}) summed over all $M_I$. The solid and dashed lines are results of the calculations (i) and (ii), respectively. Clearly, the surface environment varies the quantum interference effects, destroying the coherence of relative partial waves and changing by a few degrees the minimum of the inelastic probability distributions [Figs. \ref{Fig4}(b) and (c)]. It also affects significantly the asymptotic population of the $^{154}$Sm states and the fusion probability, as presented in Table \ref{Table}. The surface environment-induced decoherence hinders the probability flow from the elastic to the inelastic and fusion channels.           

\begin{table} \caption{Asymptotic population (\ref{eq9}) of the $^{154}$Sm states and the fusion probability for calculations including (i) only the fusion environment (Environment 1), and (ii) in addition the surface environment (Environment 2). The latter hinders the probability flow from the elastic to the inelastic and fusion channels, due to decoherence.} \label{Table}
\medskip
\begin{center}
\begin{tabular}{c|c|c}
\hline\hline &\multicolumn{1}{c}{Environment 1} &
\multicolumn{1}{|c}{Environments 1 and 2}
\\
\hline
States &~~ Population~~&~~ Population ~\\
\hline
$0^+$ &0.8376&0.8770\\
$2^+$ &0.1206&0.0961\\
$4^+$ &0.0418&0.0268\\
\hline
Fusion Probability &3.127E-3&9.106E-4\\
\hline\hline \end{tabular}
\end{center}
\end{table}

\section{Summary} 

The innovative, coupled-channels density-matrix approach to low-energy reaction dynamics has been developed further, including full angular momentum couplings and a method for extracting energy-resolved scattering information from the time-dependent density matrix. These developments have enabled a first exploration of decoherence effects on asymptotic observables, such as the angular distribution of the target excitations, and their integrated values. These are significantly affected by decoherence induced by a surface environment (related to complex, multi-nucleon/cluster transfers), which changes by a few degrees the minimum of the back-angle inelastic probability distributions and hinders the probability flow from the elastic to the inelastic and fusion channels. To test the present theory against experiments, both extensive calculations (with energy projection) and high-precision measurements of fusion and scattering observables (including those investigated here) are required. It is hoped that decoherence effects can help resolve major problems in low-energy nuclear reaction physics, such as the current inability to simultaneously describe fusion and scattering measurements of heavy ions within the standard, coupled-channels framework.          
  
\begin{acknowledgments}
The author thanks Jeff Tostevin for discussions, and Ron Johnson for constructive comments on the importance of the density-matrix off-diagonal elements (with respect to the total angular momentum) for calculating the angular distributions. Useful discussions with participants in the ECT$^*$ Workshop on Decoherence in Quantum Dynamical Systems, Trento, April 26--30, 2010, are also acknowledged. The work was supported by the UK Science and Technology Facilities Council (STFC) Grant No. ST/F012012/1.
\end{acknowledgments}


\begin{thebibliography}{99}
\bibitem{Feshbach} H. Feshbach, Ann. Phys. (N.Y.) \bf19\rm, 287 (1962), and references therein.
\bibitem{Taylor} J.R. Taylor, \emph{Scattering Theory} (John Wiley and Sons, Inc., 1972).
\bibitem{Zeh0} E. Joos, H.D. Zeh, C. Kiefer, D. Giulini, J. Kupsch and I.O. Stamatescu, 
\emph{Decoherence and the Appearance of a Classical World in Quantum Theory} (Springer, Heidelberg, 2003).
\bibitem{Zurek} W.H. Zurek, Rev. Mod. Phys. \bf75\rm, 715 (2003).
\bibitem{Schlosshauer} M. Schlosshauer, \emph{Decoherence and the Quantum-to-Classical Transition} (Springer, Berlin, 2007).
\bibitem{Breuer} H.-P. Breuer and F. Petruccione, \emph{The Theory of Open Quantum Systems} (Oxford University Press, New York, 2002).
\bibitem{Alexis0} A. Diaz-Torres, Phys. Rev. C \bf81\rm, 041603(R) (2010).
\bibitem{Saalfrank} L. Pesce and P. Saalfrank, Chem. Phys. \bf219\rm, 43 (1997); 
J. Chem. Phys. \bf108\rm, 3045 (1998).
\bibitem{Alexis1} A. Diaz-Torres et al., Phys. Rev. C \bf78\rm, 064604 (2008).
\bibitem{Alexis2} A. Diaz-Torres et al., AIP Conf. Proc. \bf1098\rm, 44 (2009).
\bibitem{Alexis3} A. Diaz-Torres, arXiv:1009.0520 [nucl-th].
\bibitem{Zeh1} H.D. Zeh, S$\acute{e}$minaire Poincar$\acute{e}$ \bf1\rm, 115 (2005).
\bibitem{Nanda0} M. Dasgupta et al., Annu. Rev. Nucl. Part. Sci. \bf48\rm, 401 (1998).
\bibitem{Hinde1} D.J. Hinde et al., Nucl. Phys. A \bf834\rm, 117c (2010), and references therein.
\bibitem{Simbel1} A.Y. Abul-Magd and M.H. Simbel, Phys. Lett. B \bf83\rm, 27 (1979).
\bibitem{Rosa1} A. De Rosa et al., Phys. Rev. C \bf40\rm, 627 (1989).
\bibitem{Balantekin1} A.B. Balantekin and N. Takigawa, Rev. Mod. Phys. \bf70\rm, 77 (1998).
\bibitem{Newton1} J.O. Newton et al., Phys. Rev. C \bf70\rm, 024605 (2004).
\bibitem{Nanda1} M. Dasgupta et al., Phys. Rev. Lett. \bf99\rm, 192701 (2007), and references therein.
\bibitem{Rae} I. Boztosun and W.D.M. Rae, Phys. Lett. B \bf518\rm, 229 (2001).  
\bibitem{Esbensen1} S. Misicu and H. Esbensen, Phys. Rev. Lett. \bf96\rm, 112701 (2006).  
\bibitem{Hagino1} T. Ichikawa, K. Hagino and A. Iwamoto, Phys. Rev. Lett. \bf103\rm, 202701 (2009).
\bibitem{Gasques1} L.R. Gasques et al., Phys. Rev. C \bf76\rm, 024612 (2007).
\bibitem{Evers1} M. Evers et al., Phys. Rev. C \bf78\rm, 034614 (2008). 
\bibitem{Lin1} C.J. Lin et al., Phys. Rev. C \bf79\rm, 064603 (2009).
\bibitem{Piasecki1} E. Piasecki et al., Phys. Rev. C \bf80\rm, 054613 (2009).
\bibitem{Yusa} S. Yusa, K. Hagino and N. Rowley, Phys. Rev. C \bf82\rm, 024606 (2010).  
\bibitem{Amos} S. Karataglidis and K. Amos, Phys. Lett. B \bf660\rm, 428 (2008). 
\bibitem{Lindblad1} G. Lindblad, Comm. Math. Phys. \bf48\rm, 119 (1976).
\bibitem{Gorini} V. Gorini et al., J. Math. Phys. \bf17\rm, 821 (1976).
\bibitem{Sandulescu1} A. Sandulescu, H. Scutaru and W. Scheid, J. Phys. A \bf20\rm, 2121 (1987); Ann. Phys. (N.Y.) \bf173\rm, 277 (1987).
\bibitem{Ronnie1} R. Kosloff, M.A. Ratner and W.B. Davis, J. Chem. Phys. \bf106\rm, 7036 (1997).
\bibitem{Saalfrank2} C. Scheurer and P. Saalfrank, J. Chem. Phys. \bf104\rm, 2869 (1996).
\bibitem{Irene} I. Burghardt, J. Phys. Chem. A \bf102\rm, 4192 (1998).
\bibitem{Bertlmann} R.A. Bertlmann et al., Phys. Rev. A \bf73\rm, 054101 (2006).
\bibitem{Erratum} Expression (6) in Ref. \cite{Alexis1} is incorrect, but the correction of this error does not affect any conclusions of that paper.
\bibitem{Saalfrank3} C. Scheurer and P. Saalfrank, Chem. Phys. Lett. \bf245\rm, 201 (1995).
\bibitem{Matsuo} M. Matsuo et al., Nucl. Phys. A \bf649\rm, 379c (1999).
\bibitem{Guo} L. Guo et al., Nucl. Phys. A \bf753\rm, 136 (2005).
\bibitem{Ring} E. Litvinova et al., Phys. Rev. C \bf75\rm, 064308 (2007).
\bibitem{McIntosh} A.B. McIntosh et al., Phys. Rev. Lett. \bf99\rm, 132701 (2007).
\bibitem{Schafer} K.J. Schafer and K.C. Kulander, Phys. Rev. A \bf42\rm, 5794 (1990); 
Comp. Phys. Comm. \bf63\rm, 427 (1991).
\bibitem{Tannor} D.J. Tannor, \emph{Introduction to Quantum Mechanics: A Time-Dependent Perspective} (University Science Books, Saulito, 2007).
\bibitem{Yabana} K. Yabana, Prog. Theor. Phys. \bf97\rm, 437 (1997).
\bibitem{CCFULL} K. Hagino, N. Rowley and A.T. Kruppa, Comp. Phys. Comm. \bf123\rm, 143 (1999).
\bibitem{Faber} W. Huisinga et al., J. Chem. Phys. \bf110\rm, 5538 (1999). 
\bibitem{Kosloff2} R. Kosloff, Annu. Rev. Phys. Chem. \bf45\rm, 145 (1994).
\bibitem{KR} M.W. Kermode and N. Rowley, Phys. Rev. C \bf48\rm, 2326 (1993).
\bibitem{Satchler} G.R. Satchler, \emph{Direct Nuclear Reactions} (Oxford University Press, Oxford, 1983) p. 684.
\bibitem{Klum} K. Blum, \emph{Density Matrix Theory and Applications} (Second Edition, Plenum Press, New York, 1996) p. 39.

\end{thebibliography}
\end{document}